\documentclass{article}

\usepackage{amsfonts}
\usepackage{amsmath}
\usepackage{amssymb}
\usepackage{graphicx}
\usepackage{epsfig}
\usepackage{amscd}
\usepackage{color}
\usepackage{amsthm}

\newtheorem{theor}{Theorem}
\newtheorem{defi}{Definition}
\newtheorem{prop}{Proposition}
\newtheorem{corol}{Corollary}

\begin{document}

\begin{center}
\Large{\textbf{Another dual formulation of the separability problem}}
\end{center}

\begin{center}
D. Salgado \& J.L. S\'{a}nchez-G\'{o}mez\\ 
Dpto.  F\'{\i}sica Te\'{o}rica, Universidad Aut\'{o}noma de Madrid, Spain \\
\texttt{david.salgado@uam.es} \& \texttt{jl.sanchezgomez@uam.es} \\

\medskip

M.\ Ferrero\\
Dpto.\ F\'{\i}sica, Universidad de Oviedo, Spain\\
\texttt{maferrero@uniovi.es}

\medskip
\end{center}

\vspace{1cm}

\begin{center}
\centerline{\textbf{Abstract}}
\begin{minipage}{10cm} 
We show how the separability problem is dual to that of decomposing any given matrix into a conic combination of rank-one partial isometries, thus offering a duality approach different to the positive maps characterization problem. Several inmediate consequences are analyzed: (i) a sufficient criterion for separability for bipartite quantum systems, (ii) a complete solution to the separability problem for pure states also of bipartite systems independent of the classical Schmidt decomposition method and (iii) a natural generalization of these results to multipartite systems.
\end{minipage} 
\end{center}

\vspace{1cm} 

\section{Introduction}
\label{Intro}
Entanglement is at the heart of the most surprising quantum phenomena, especially those sustaining the quantum processing of information (cf.\ e.g.\ \cite{ChuNie00a,BouEkeZei00a}). However, a complete characterization of entanglement of compound quantum systems is not nowadays at hand. There exists a great deal of either sufficient or necessary conditions for a system to show entanglement (cf.\ \cite{LewBruCirKraKusSamSanTar00a,BruCirHorHulKraLewSan01a,HorHorHor01a,Bru02a,Ter02a} and multiple references therein), and those both necessary and sufficient criteria either apply to a limited number of systems \cite{HorHorHor96a,HorHor99a} or lack of an operational (user-friendly) character \cite{Ter00a,KraCirKarLew00a,Rud00a,DohParSpe04a}. Among all of them, the most celebrated is the well-known PPT or Peres-Horodecki criterion \cite{Per96a,HorHorHor96a}, which discerns between entangled and separable states just by checking the positivity of the density matrix of the bipartite compound system after performing the partial transposition \footnote{Given a matrix $Q\in\mathcal{M}_{m}\otimes\mathcal{M}_{n}$, which can always be written as $Q=\sum_{ij=1}^{m}E_{ij}\otimes Q_{ij}$, where $E_{ij}$ denotes the Weyl matrices, the partial transposition with respect to the first factor is defined as $Q^{T_{1}}=\sum_{ij=1}^{m}E_{ij}^{T}\otimes Q_{ij}$ and with respect to the second factor as $Q^{T_{2}}=\sum_{ij=1}^{m}E_{ij}\otimes Q_{ij}^{T}$, where $A^{T}$, as usual, denotes the transpose of $A$.}. The grandeur of the PPT criterion resides on the translation of the separability question to a dual problem, namely, the characterization of positive linear maps between matrix algebras \cite{HorHorHor96a}. Since this long-standing mathematical problem has only been solved for maps between two two-dimensional \cite{Sto63a} and a two-dimensional and a three-dimensional matrix algebra \cite{Wor76a}, this criterion only works for $2\times 2$ and $2\times 3$ systems. The ultimate reason for this limitation then turns out to be the lack of knowledge about the structure of the set of positive linear maps. In the multipartite case, the situation is even more dramatic, since no full characterization is known.\\

Here we formulate an equivalent dual problem to that of separability of quantum states by resorting to two mathematical tools commonly used in Matrix Analysis (cf.\ e.g.\ \cite{HorJoh91a}), namely the $vec$ operator carrying matrices onto vectors and the \emph{tilde} transform, realigning matrix elements into another matrix with the same total number of elements, although in general, different dimensions. As a direct consequence we obtain both a sufficient condition for separability and a characterization of product states for bipartite and multipartite systems of any dimension, thus offering an alternative to the classical Schmidt decomposition in the case of pure states. Very close objects have already been used in this context in the so-called \emph{realignment method} \cite{CheWu03a}, although the use of a (Ky-Fan) norm in this approach drives us only to a necessary condition for separability (cf.\ \cite{Rud02a,Rud03a} for an equivalent approach and \cite{CheWu02a} for the generalization of this method to multipartite systems.).\\

We report our results as follows. In section \ref{VecTil} we introduce the mathematical tools, namely the $vec$ operator and the \emph{tilde} transform; in section \ref{DuaForSomCon} we establish the new duality problem and analyse some of its first consequences for bipartite systems of any dimensions; in section \ref{MulCas} we extend preceding results to the multipartirte case; we illustrate the results with some examples in section \ref{Exa}; we conclude in section \ref{Con} with some comments and remarks.

\section{The $vec$ operator and the \emph{tilde} transform}
\label{VecTil}
Throughout the whole paper we will denote the set of $n$-dimensional matrices by $\mathcal{M}_{n}$ and vectors of any dimension by the usual \emph{bra-ket} notation. The $vec$ operator is commonly defined as a linear isomorphism carrying matrices $M$ of arbitrary dimensions $m\times n$ to $mn-$dimensional vectors obtained by stacking the matrix columns successively from left to right (cf.\ \cite{HenSea81a} for a mathematical review):

\begin{defi}
Let $Q=[q_{1} \dots q_{n}]\in\mathcal{M}_{n}$, where $q_{i}$ denotes the $i$th column of $Q$. Then the $vec$ operator is defined as 

\begin{eqnarray}\label{DefVec}
vec:\mathcal{M}_{n}&\to&\mathbb{C}^{n^{2}}\nonumber\\
Q&\to&|vec(Q)\rangle\equiv\left(\begin{array}{c}
q_{1}\\
\vdots\\
q_{n}
\end{array}\right)
\end{eqnarray} 
\end{defi}

 This stacking is conventional and generalizations of this operator can be found in the literature \cite{KonNeuWan91a} with different purposes and applications. The generalization to rectangular matrices is elementary, although useless for our immediate purposes. It is straightforward to prove the following

\begin{prop}
The $vec$ operator is a linear isomorphism.
\end{prop}

Related to it, one can also introduce a so-called \emph{tilde} transform \cite{KonNeuWan91a}, which for the same reasons as before we will only define for square matrices (the generalization to rectangular matrices is also elementary):

\begin{defi}
Let $Q\in\mathcal{M}_{m}\otimes\mathcal{M}_{n}$, which can always be written as $Q=\sum_{ij=1}^{m}E_{ij}\otimes Q_{ij}$, where $E_{ij}$ are the $m-$dimensional square Weyl matrices. The \emph{tilde} transform of $Q$ is defined as 

\begin{equation}
\tilde{Q}\equiv t(Q)=\sum_{ij=1}^{m}|vecQ_{ij}\rangle\langle vecE_{ij}^{\dagger}|
\end{equation}
\end{defi}

Notice that $\tilde{Q}$ is an $n^{2}\times m^{2}$ rectangular matrix. This definition drives us to a realigned matrix different from that used in \cite{CheWu03a} to establish a new (necessary) separability criterion, namely, the \emph{realignment criterion}. It is straightforward to convince oneself that the \emph{realignment method} can also be established with the above convention, since the Ky-Fan norm properties \cite{HorJoh91a} of the \emph{tilde}-transformed matrices are equivalent. Let us provide the following illustrative example in $\mathcal{M}_{2}\otimes\mathcal{M}_{2}$:

\begin{equation}\label{ExaTil}
Q=\begin{pmatrix}
q_{11}&q_{12}&q_{13}&q_{14}\\
q_{21}&q_{22}&q_{23}&q_{24}\\
q_{31}&q_{32}&q_{33}&q_{34}\\
q_{41}&q_{42}&q_{43}&q_{44}
\end{pmatrix}\Rightarrow t(Q)=\begin{pmatrix}
q_{11}&q_{31}&q_{13}&q_{33}\\
q_{21}&q_{41}&q_{23}&q_{43}\\
q_{12}&q_{32}&q_{14}&q_{34}\\
q_{22}&q_{42}&q_{24}&q_{44}
\end{pmatrix}
\end{equation} 

The main property why this definition has been adopted is the following

\begin{prop}\label{ProTil}
Let $Q_{1}\otimes Q_{2}\in\mathcal{M}_{m}\otimes\mathcal{M}_{n}$. Then $t(Q_{1}\otimes Q_{2})=|vecQ_{2}\rangle\langle vecQ_{1}^{\dagger}|$.
\end{prop}

\begin{proof}
The proof follows inmediately from the definition of the $tilde$ transform and the linear properties of the $vec$ operator:

\begin{subequations}
\begin{eqnarray}
t(Q_{1}\otimes Q_{2})&=&\sum_{ij=1}^{m}t(q_{ij}^{(1)}E_{ij}\otimes Q_{2})=\\
&=&\sum_{ij=1}^{m}q_{ij}^{(1)}|vecQ_{2}\rangle\langle vecE_{ji}|=\\
&=&|vecQ_{2}\rangle\langle vec(\sum_{ij=1}^{m}q_{ij}^{(1)*}E_{ji})|=\\
&=&|vecQ_{2}\rangle\langle vecQ_{1}^{\dagger}|
\end{eqnarray}
\end{subequations}
\end{proof}

As before, it is elementary to prove the following

\begin{prop}
The \emph{tilde} transform is a linear isomorphism.
\end{prop}

 Notice that in Quantum Mechanics we are mostly involved with Hermitian matrices, thus for any product state $\rho_{1}\otimes\rho_{2}$, the latter proposition reduces to $t(\rho_{1}\otimes\rho_{2})=|vec\rho_{2}\rangle\langle vec\rho_{1}|$. Had we used the matrix realigment criterion of \cite{CheWu03a}, we would have ended with the property $t(A\otimes B)=|vec A\rangle\langle vec B^{*}|$, which introduces the nuisance of taking complex conjugates of all elements of $B$. In table \ref{DefsTilTra} we include all other possible definitions of a \emph{tilde} transform driving to a similar tensor product property. We have chosen definition 8 in order to have either $vec E_{ij}$ in the \emph{bra} part of its definition (so that the realingment does not involve also complex conjugation) and the adjoint of one of the factors in the tensor product property (since this operation will not affect Hermitian matrices and, in particular, density matrices). Any of these definitions allows us to reproduce the subsequent results in a similar fashion.\\

\begin{table}[htp]
\begin{tabular}{c||c}
Definitions of $t$&Tensor product property\\\hline\hline\\
1. $t(Q)=\sum_{ij=1}^{m}|vec E_{ij}\rangle\langle vecQ_{ij}^{*}|$\hspace*{2mm} &\hspace*{2mm}$t(Q_{1}\otimes Q_{2})=|vec Q_{1}\rangle\langle vec Q_{2}^{*}|$\\
2. $t(Q)=\sum_{ij=1}^{m}|vec E_{ij}\rangle\langle vecQ_{ij}|$\hspace*{2mm}&\hspace*{2mm}$t(Q_{1}\otimes Q_{2})=|vec Q_{1}^{*}\rangle\langle vec Q_{2}|$\\
3. $t(Q)=\sum_{ij=1}^{m}|vec E_{ij}\rangle\langle vecQ_{ij}^{\dagger}|$\hspace*{2mm}&\hspace*{2mm}$t(Q_{1}\otimes Q_{2})=|vec Q_{1}\rangle\langle vec Q_{2}^{\dagger}|$\\
4. $t(Q)=\sum_{ij=1}^{m}|vec E_{ji}\rangle\langle vecQ_{ij}|$\hspace*{2mm}&\hspace*{2mm}$t(Q_{1}\otimes Q_{2})=|vec Q_{1}^{\dagger}\rangle\langle vec Q_{2}|$\\
5. $t(Q)=\sum_{ij=1}^{m}|vec Q_{ij}^{*}\rangle\langle vecE_{ij}|$\hspace*{2mm}&\hspace*{2mm}$t(Q_{1}\otimes Q_{2})=|vec Q_{2}^{*}\rangle\langle vec Q_{1}|$\\
6. $t(Q)=\sum_{ij=1}^{m}|vec Q_{ij}\rangle\langle vecE_{ij}|$\hspace*{2mm}&\hspace*{2mm}$t(Q_{1}\otimes Q_{2})=|vec Q_{2}\rangle\langle vec Q_{1}^{*}|$\\
7. $t(Q)=\sum_{ij=1}^{m}|vec Q_{ij}^{\dagger}\rangle\langle vecE_{ij}|$\hspace*{2mm}&\hspace*{2mm}$t(Q_{1}\otimes Q_{2})=|vec Q_{2}^{\dagger}\rangle\langle vec Q_{1}|$\\
8. $t(Q)=\sum_{ij=1}^{m}|vec Q_{ij}\rangle\langle vecE_{ji}|$\hspace*{2mm}&\hspace*{2mm}$t(Q_{1}\otimes Q_{2})=|vec Q_{2}\rangle\langle vec Q_{1}^{\dagger}|$\\
\end{tabular}
\caption{\label{DefsTilTra}Definitions of the \emph{tilde} transform with the corresponding tensor product property.}
\end{table}

 The matrices of the form $|u\rangle\langle v|$ belong to a special class of matrices called \emph{partial isometries}, thus we will refer to them as rank-one partial isometries \cite{HorJoh91a}. Also, we will give $vec^{-1}$, the inverse of the $vec$ operator, carrying vectors to matrices, a special name: $mat\equiv vec^{-1}$. Its action is clear from that of $vec$ (read relation \eqref{DefVec} from right to left). Note that applying the \emph{tilde} transform twice to a matrix leaves it invariant, i.e.\ the \emph{tilde} transform is its own inverse (check for instance equation \eqref{ExaTil}).\\

Finally, let us comment that under mathematical rigor we should have denoted the $vec$ operator as $vec_{n}$ and the $mat$ operator as $mat_{n}$, keeping clear upon which matrix algebra they are operating. This is also valid for the \emph{tilde} transform, which should have been denoted as $t_{m,n}$. However in order to ease the notation, we will assume that these matrix vector spaces are fixed from the beginning so that the corresponding dimensions are known.

\section{The dual formulation and some consequences}
\label{DuaForSomCon}

We formulate our main result for bipartite systems, which establish the dual formulation of the separability problem:

\begin{theor}\label{MaiRes} Let $\rho$ be a density matrix of an $m\times n$ bipartite system. Then $\rho$ is separable if, and only if, $\tilde{\rho}$ admits a conic decomposition into rank-one partial isometries, i.e. 

\begin{equation}
\tilde{\rho}=\sum_{i=1}^{K}\mu_{i}|u_{i}\rangle\langle v_{i}|\quad \mu_{i}> 0
\end{equation}  

\noindent where $mat|u_{i}\rangle\geq 0$ and $mat|v_{i}\rangle\geq 0$ for all $i=1,\dots,K$.
\end{theor}
\begin{proof}
If $\rho$ is separable, then there exist positive matrices $\rho_{i}\in\mathcal{M}_{m}$ and $\sigma_{i}\in\mathcal{M}_{n}$ (density matrices, indeed) and positive numbers $\mu_{i}$ such that $\rho=\sum_{i=1}^{K}\mu_{i}\rho_{i}\otimes\sigma_{i}$. Applying the $tilde$ transform upon $\rho$, we obtain
\begin{subequations}
\begin{eqnarray}
\tilde{\rho}&=&\sum_{i=1}^{K}\mu_{i}t(\rho_{i}\otimes\sigma_{i})=\\
&=&\sum_{i=1}^{K}\mu_{i}|vec\sigma_{i}\rangle\langle vec\rho_{i}^{\dagger}|=\\
&=&\sum_{i=1}^{K}\mu_{i}|vec\sigma_{i}\rangle\langle vec\rho_{i}|\equiv\\
&\equiv&\sum_{i=1}^{K}\mu_{i}|u_{i}\rangle\langle v_i |
\end{eqnarray}
\end{subequations}

\noindent where clearly $mat|u_{i}\rangle=\sigma_{i}\geq 0$ and $mat|v_{i}\rangle=\rho_{i}\geq 0$.\\
Conversely, if such a decomposition exists, then
\vspace*{-0.2cm}
\begin{subequations}
\begin{eqnarray}
\rho&=&t(\tilde{\rho})\equiv\tilde{\tilde{\rho}}=\\
&=&\sum_{i=1}^{K}\mu_{i}t(|u_{i}\rangle\langle v_{i}|)=\\
&=&\sum_{i=1}^{K}\mu_{i}(mat|v_{i}\rangle)^{\dagger}\otimes mat|u_{i}\rangle=\\
&=&\sum_{i=1}^{K}\mu_{i} mat|v_{i}\rangle\otimes mat|u_{i}\rangle
\end{eqnarray}
\end{subequations}

\noindent Thus, since $mat|u_{i}\rangle$ and $mat|v_{i}\rangle$ are (Hermitian) positive matrices, $\rho$ is separable. 
\end{proof}

As a first consequence of this result, the singular value decomposition \cite{HorJoh91a} provides a method to find a conic decomposition into partial isometries, thus furnishing a sufficient criterion of separability:

\begin{corol}\label{NecSepCri}
Let $\rho$ be a density matrix of an $m\times n$ bipartite system. Let $\{|v_{i}\rangle\}_{i=1,\dots,q}$ and $\{|w_{i}\rangle\}_{i=1,\dots,q}$ be left and right singular vectors, respectively, associated to non-null singular values of $\tilde{\rho}$. If $mat |v_{i}\rangle\geq 0$ and $mat|w_{i}\rangle\geq 0$ for all $i=1,\dots,q$, then $\rho$ is separable.
\end{corol}

\begin{proof}
Let us recall \cite{HorJoh91a} that the singular value decomposition of a matrix $Q=V\Sigma W^{\dagger}$ can be rewritten as $Q=\sum_{i=1}^{q}\sigma_{i}|v_i\rangle\langle w_{i}|$, where $|v_{i}\rangle$ and $|w_{i}\rangle$ are left and right singular vectors associated to non-null singular values $\sigma_{i}$. The rest follows elementary from theorem \ref{MaiRes}.
\end{proof}

The necessity is unattainable in general, since singular vectors are always orthonormal, i.e. $\langle v_{i}|v_{j}\rangle=\langle w_{i}|w_{j}\rangle=\delta_{ij}$. This restriction is, however, absent from theorem \ref{MaiRes}, thus the singular value decomposition cannot provide the required generality. Nevertheless, a notable advantage of this sufficient criterion is that, in the positive case, it yields by construction a convex combination of $\rho$ in product states.\\

Another inmediate consequence of our result is a necessary and sufficient criterion to know whether a given state is a product state or not:

\begin{corol}\label{BipProSta}
Let $\rho$ be a density matrix of a bipartite system. Then $\rho=\rho_{1}\otimes\rho_{2}$ if, and only if, $rank(\tilde{\rho})=1$ and the unique left and right singular vectors $|u\rangle$ and $|v\rangle$ associated to the non-null singular value of $\tilde{\rho}$ satisfy $mat |u\rangle\geq 0$, $mat |v\rangle\geq 0$.
\end{corol}

Notice that this result offers a separability check for pure states alternative to the common Schmidt decomposition \cite{Sch07a,EkeKni95a}, which, in addition, in the positive case also provides the corresponding factors. Starting from a pure state $|\Psi\rangle$, all we have to do is to apply corollary \ref{BipProSta} to $\rho_{\Psi}=|\Psi\rangle\langle\Psi|$, together with the further conditions $r(mat|u\rangle)=r(mat|v\rangle)=1$. Moreover, as shown below, this method can also be exported to the multipartite case in a natural way.\\

Another characterization of those relevant rank-one partial isometries fulfilling the conditions of theorem \ref{MaiRes} can be obtained by resorting to the structure of the cone of positive matrices \cite{HilWat87a}: $Q\geq 0$ if, and only if, there exist positive numbers $\lambda_{i}$ and orthogonal projectors $P_{u_{i}}\equiv|u_{i}\rangle\langle u_{i}|$ such that $Q=\sum_{i}\lambda_{i} P_{u_{i}}$. Keeping in mind that $vec P_{u}=|u^{*}\rangle\otimes|u\rangle$, then theorem \ref{MaiRes} can be reformulated as
\begin{theor}\label{RefMaiRes}
 Let $\rho$ be a density matrix of an $m\times n$ bipartite system. Then $\rho$ is separable if, and only if, $\tilde{\rho}$ admits a conic decomposition into rank-one partial isometries, i.e. 
\begin{equation}
\tilde{\rho}=\sum_{i=1}^{K}\mu_{i}|u_{i}\rangle\langle v_{i}|\quad \mu_{i}> 0
\end{equation}  

\noindent where for each $i=1,\dots,K$
\begin{subequations}
\begin{eqnarray}
|u_{i}\rangle&=&|x_{i}^{*}\rangle\otimes|x_{i}\rangle\\
|v_{i}\rangle&=&|y_{i}^{*}\rangle\otimes|y_{i}\rangle 
\end{eqnarray}
\end{subequations}
\end{theor}

\begin{proof}
Starting from $\tilde{\rho}=\sum_{i}\lambda_{i}|u_{i}\rangle\langle v_{i}|$, where $\lambda_{i}>0$ and $mat|u_{i}\rangle,mat|v_{i}\rangle\geq 0$, insert $|u_{i}\rangle=\sum_{k}\alpha_{k}^{(i)}|x_{k}^{(i)*}\rangle\otimes|x_{k}^{(i)}\rangle$ and $|v_{i}\rangle=\sum_{q}\beta_{q}^{(i)}|y_{q}^{(i)*}\rangle\otimes|y_{q}^{(i)}\rangle$ to obtain

\begin{equation}
\tilde{\rho}=\sum_{ikq}\lambda_{i}\alpha_{k}^{(i)}\beta_{q}^{(i)}\left(|x_{k}^{(i)*}\rangle\otimes|x_{k}^{(i)}\rangle\right)\left(\langle y_{q}^{(i)*}|\otimes\langle y_{q}^{(i)}|\right)
\end{equation}

Merge the enumerable indices into a single one to arrive at the desired result. 

\end{proof}

Regretfully this result does not either provide us with an operational separability check, since an algorithm to find such a decomposition into the preceding rank-one partial isometries is in general unknown. However, a slight difference with respect to the positive maps characterization problem is detected. Let us make use of the language of the theory of cones \cite{Bar81a} \footnote{A cone $K$ is a closed set $K$ of a vector space $V$ such that $\lambda K\subset K$ for all $\lambda\geq 0$; in the theory of cones referred to above a cone is always (i) convex ($\lambda K+(1-\lambda)K\subset K$), (ii) pointed $K\cap (-K)=\{0\}$ and (iii) reproducing $K-K=V$. One of the most famous cones is that of positive semidefinite matrices (cf.\ \cite{HilWat87a}). The set of cone-preserving linear maps can be also proved to be a cone. Among the most relevant properties of a cone is that any of its element can be expressed as a linear combination of its extremal elements with positive coefficients (conic combination).}. The positive maps characterization problem can be formulated in terms of the analysis of the structure of the cone $\Pi_{m,n}$ of positive maps between an $m$- and an $n$-dimensional matrix algebra mapping positive matrices of $\mathcal{M}_{m}$ into positive matrices of $\mathcal{M}_{n}$: the set of extremal elements of $\Pi_{m,n}$, denoted as $\textrm{Ext}\Pi_{m,n}$, is only known in the cases $(m,n)=(2,2)$, $(2,3)$ and $(3,2)$. The reformulation proposed here reduces the separability problem to find a criterion to discern whether a given arbitrary $n^{2}\times m^{2}$ rectangular matrix belongs to the cone $\tilde{K}_{n^{2},m^{2}}$, whose set of extremal elements is, by definition, $\textrm{Ext}\tilde{K}_{n^{2},m^{2}}\equiv\{|u\rangle\langle v|:|u\rangle=|x^{*}\rangle\otimes|x\rangle\in\mathbb{C}^{n}\otimes\mathbb{C}^{n}, |v\rangle=|y^{*}\rangle\otimes|y\rangle\in\mathbb{C}^{m}\otimes\mathbb{C}^{m}\}$. Now the set of extremal elements is known, thus everything is reduced to find an operational criterion to decide whether $\tilde{\rho}$ belongs to $K_{n^{2},m^{2}}$ or not.\\

\section{The multipartite case}
\label{MulCas}
To analyse the multipartite case, we need to generalize the $tilde$ transform. Let us first recall that an arbitrary matrix $Q\in\mathcal{M}_{n_{1}}\otimes\dots\otimes\mathcal{M}_{n_{m}}$ can be written in any of these $m$ forms:
\begin{subequations}
\begin{eqnarray}
Q&=&\sum_{i_2,j_2=1}^{n_2}\dots\sum_{i_{m},j_{m}=1}^{n_{m}}Q^{(1)}_{i_{2}j_{2}\dots i_{m}j_{m}}\otimes E_{i_{2}j_{2}}\otimes\dots\otimes E_{i_{m}j_{m}}\\
&=&\sum_{i_1,j_1=1}^{n_1}\sum_{i_{3},j_{3}=1}^{n_{3}}\dots\sum_{i_{m},j_{m}=1}^{n_{m}} E_{i_{1}j_{1}}\otimes Q^{(2)}_{i_{1}j_{1} i_{3}j_{3}\dots i_{m}j_{m}}\otimes\dots\otimes E_{i_{m}j_{m}}\\
&=&\hspace*{3cm}\vdots\nonumber\\
&=&\sum_{i_1,j_1=1}^{n_1}\dots\sum_{i_{n_{m}-1},j_{n_{m}-1}=1}^{n_{m}-1} E_{i_{1}j_{1}}\otimes\dots\otimes E_{i_{m-1}j_{m-1}}\otimes Q^{(m)}_{i_{1}j_{1}\dots i_{m-1}j_{m-1}}\nonumber\\
\end{eqnarray}
\end{subequations}

\noindent where $E_{i_{p}j_{p}}$ denotes the Weyl matrices in the space $\mathcal{M}_{n_{p}}$. The matrices $Q^{(k)}\equiv Q_{i_{1}j_{1}\dots}^{(k)}$ have dimension $n_{k}\times n_{k}$. A systematic procedure to find these matrices $Q^{(k)}$ arises from the use of the vec-permutation matrix $P(m,n)\equiv\sum_{i=1}^{m}\sum_{j=1}^{n}E_{ij}\otimes E_{ji}$ and the property $B\otimes A=P^{T}(m,n)(A\otimes B)P(m,n)$, where $A\in\mathcal{M}_{m}$ and $B\in\mathcal{M}_{n}$ \cite{HorJoh91a}.\\

With the same spirit as before we introduce the generalized $tilde$ transforms:

\begin{defi}
Let $Q\in\mathcal{M}_{n_{1}}\otimes\dots\otimes\mathcal{M}_{n_{m}}$. The $k$th tilde transform of $Q$ is defined as


\begin{equation}
\hspace*{-2.5cm}t_{k}(Q)=\sum_{i_{1}j_{1}=1}^{n_{1}}\hspace*{-1mm}\dots\hspace*{-1mm}\sum_{i_{k-1}j_{k-1}=1}^{n_{k-1}}\sum_{i_{k+1}j_{k+1}=1}^{n_{k+1}}\hspace*{-1mm}\dots\hspace*{-1mm}\sum_{i_{m}j_{m}=1}^{n_{m}}|vecQ^{(k)}_{i_{1}j_{1}\dots i_{i_{m}j_{m}}}\rangle\langle vec (E_{i_{1}j_{1}}^{\dagger}\otimes \dots\otimes E_{i_{k-1} j_{k-1}}^{\dagger}\otimes E_{i_{k+1} j_{k+1}}^{\dagger}\otimes\dots \otimes E_{i_{m} j_{m}}^{\dagger})|
\end{equation}

\end{defi}

This generalized $tilde$ transform also satisfies a similar property:
\begin{prop}
Let $Q\in\mathcal{M}_{n_{1}}\otimes\dots\otimes\mathcal{M}_{n_{m}}$. Then
\begin{equation}
t_{k}(\bigotimes_{i=1}^{m}Q_{i})=|vec Q_{k}\rangle\langle vec(Q_{1}^{\dagger}\otimes\dots\otimes Q_{k-1}^{\dagger}\otimes Q_{k+1}^{\dagger}\otimes\dots\otimes Q_{m}^{\dagger}|
\end{equation}
\end{prop}
\begin{proof}
It is an inmediate generalization of the proof of proposition \ref{ProTil}.
\end{proof}
The analysis of the separability properties of a multipartite state depends on the particular partition chosen for the compound system. Here we will focus on full separability, i.e. on the conditions upon which a state $\rho$ of an $m-$partite system can be written as $\rho=\sum_{i_{1}\dots i_{m}}\lambda_{i_{1}\dots i_{m}}\rho_{i_{1}}\otimes\dots\otimes\rho_{i_{m}}$.
\begin{theor}
Let $\rho$ be a state of an $n_{1}\times\dots\times n_{m}$ $m-$partite system. Then $\rho$ is fully separable if, and only if, $t_{k}(\rho)$ admits a conic decomposition into rank-one partial isometries
\begin{equation}
t_{k}(\rho)=\sum_{i=1}^{q_{k}}\mu_{i}^{(k)}|u_{i}^{(k)}\rangle\langle v_{i}^{(k)}|
\end{equation}
\noindent where $mat|u_{i}^{(k)}\rangle\geq 0$ and $mat|v_{i}^{(k)}\rangle\geq 0$ for all $i=1,\dots,q_{k}$ and for all $k=1,\dots,m$.

\end{theor}
The proof is a tedious, although elementary, algebraic generalization of that of theorem \ref{MaiRes}. The preceding corolaries can also be adapted to the multipartite case, in particular, the use of the singular value decomposition also allows us to define the following sufficient criterion of full separability:
\begin{corol}
Let $\rho$ be a state of an $n_{1}\times\dots\times n_{m}$ $m-$partite system. Let $\{|v_{i}^{(k)}\rangle\}_{i=1,\dots,q_{k}}$ and $\{|w_{i}^{(k)}\rangle\}_{i=1,\dots,q_{k}}$ be left and right singular vectors, respectively, associated to non-null singular values of $t_{k}(\rho)$. If $mat|v_{i}^{(k)}\rangle\geq 0$ and $mat|w_{i}^{(k)}\rangle\geq 0$ for all $i=1,\dots,q_{k}$ and for each $k=1,\dots,m$, then $\rho$ is fully separable.
\end{corol}
Once more, the orthonormality of singular vectors prevents us from attaining the necessity of this criterion. As above, we can also establish a necessary and sufficient test of separability for pure states, thus circunvemting the lack of generalization of the Schmidt decomposition to multipartite systems \cite{Per95a}.
\begin{corol}
Let $|\Psi\rangle$ be a pure state of an $n_{1}\times\dots\times n_{m}$ $m-$partite system. Then $|\Psi\rangle$ is a product state (fully separable) if, and only if, $rank(t_{k}(\rho_{\Psi}))=1$ for all $k=1,\dots,m$ and the unique left and right singular vectors $|v^{(k)}\rangle$  and $|w^{(k)}\rangle$ corresponding to the non-null singular value of $t_{k}(\rho_{\Psi})$ satisfy $mat|v^{(k)}\rangle\geq 0$, $mat|w^{(k)}\rangle\geq 0$ and $r(mat|v^{(k)}\rangle)=r(mat|w^{(k)}\rangle)=1$ for each $k=1,\dots,m$.
\end{corol}
The analysis of other partitions can be easily undertaken with the following recipe. For concreteness' sake, let us suppose we want to investigate whether a given $5-$partite system pure state $|\Psi\rangle$ can be decomposed as $|\Psi\rangle=|\psi\rangle_{12}\otimes|\phi\rangle_{3}\otimes|\varphi\rangle_{45}$, i.e.\ whether it admits a $(12)(3)(45)$ separable partition. Then we treat systems $1$ and $2$ as an $n_{1}n_{2}$ single system, as well as systems $4$ and $5$ as an $n_{4}n_{5}$ single system, and then apply the preceding result. The analysis of the separability of the factors $|\psi\rangle_{12}$ and $|\psi\rangle_{45}$ can also be undertaken with the same tools, thus providing a systematic method to detect multiseparability. These results complement those of \cite{Tha99a,AciAndCosJanLatTar00a}.

\section{Examples}
\label{Exa}
As a first elementary and well-known example we will prove how the completely depolarized state in an $m\times n$ system is separable:

\begin{equation}
\rho=\frac{1}{mn}\mathbb{I}_{mn}=\frac{1}{mn}\sum_{ij}^{m}E_{ij}^{(m)}\otimes\delta_{ij}\mathbb{I}_{n}\Rightarrow\tilde{\rho}=\frac{1}{mn}|vec\mathbb{I}_{n}\rangle\langle vec\mathbb{I}_{m}|
\end{equation}

Since $mat|vec\mathbb{I}_{k}\rangle\geq 0$ for all $k=1,2,\dots$, $\rho$ is separable.\\

A second example is provided with a Bell basis state, which serves to illustrate that if a state is entangled, then it cannot fulfill the conditions of corollary \ref{NecSepCri}. We firstly need the following elementary property of $mat$:

\begin{prop}
The $mat$ operator is an isometry between $\mathbb{C}^{n^{2}}$ and $\mathcal{M}_{n}$, where the inner products are the standard complex scalar product $\langle u|v\rangle=\sum_{k}u_{k}^{*}v_{k}$ and the trace scalar product $(A|B)=\textrm{tr}(A^{\dagger}B)$. 
\end{prop}

Then, it is easy to convince oneself that the singular value of any Bell basis state is $1$ (fourfold), and since it is impossible to find four pairwise orthonormal vectors $v_i\in\mathbb{C}^{4}$ such that $mat v_{i}\geq 0$ and $V=[v_{1}\ v_{2}\ v_{3}\ v_{4}]$ is unitary \footnote{This follows from the preceding proposition and the fact that is is impossible to find four two-dimensional pairwise orthonormal positive matrices.}, we can conclude that the conditions of corollary \ref{NecSepCri} are impossible to fulfill.\\

Another example illustrating how the singular value decomposition fails to solve the separability problem is given by the state $\rho=\frac{1}{2}\left(P_{e_{1}}\otimes P_{e_{1}}+P_{x}\otimes P_{x}\right)$, where $|x\rangle=\frac{1}{\sqrt{2}}\left(|e_{1}\rangle+|e_{2}\rangle\right)$. The singular value decomposition of $\tilde{\rho}$ is given by 

\begin{equation}
\tilde{\rho}=\frac{3}{4}|u_{1}\rangle\langle u_{1}|+\frac{1}{4}|u_{2}\rangle\langle u_{2}|
\end{equation}

\noindent where $|u_{1}\rangle=\left(\begin{smallmatrix}
\frac{\sqrt{3}}{2}\\
\frac{1}{2\sqrt{3}}\\
\frac{1}{2\sqrt{3}}\\
\frac{1}{2\sqrt{3}}\\
\end{smallmatrix}\right)$ and  $|u_{2}\rangle=\left(\begin{smallmatrix}
\frac{1}{2}\\
-\frac{1}{2}\\
-\frac{1}{2}\\
-\frac{1}{2}\\
\end{smallmatrix}\right)$. Clearly $mat|u_{2}\rangle\ngeq 0$, thus we cannot conclude the separability of $\rho$. This illustrates the need to find the decomposition of $\tilde{\rho}$ into rank-one partial isometries without the orthogonality relation.


\section{Conclusions}
\label{Con}
In summary, we have reformulate the separability problem of quantum states as a descomposition into partial isometries of the density matrix transformed under the $tilde$ transform. We have deduced several facts from this reformulation: (i) given that both the $tilde$ transform and the singular value decomposition of an arbitrary matrix are computationally accesible tasks, we have provided a systematic sufficient test of separability both for bipartite and multipartite systems of any dimension, and (ii) we have also provided a complete solution independent of the Schmidt decomposition for the pure-state case in both bipartite and multipartite systems.\\

Nevertheless there is still a relevant  open question: a criterion to discern whether a given matrix admits a decomposition into rank-one partial isometries fulfilling the conditions of theorem \ref{MaiRes} or of theorem \ref{RefMaiRes}. This demands a systematic study of the cone of matrices whose extremals are of the form $|u\rangle\langle v|$, with $|u\rangle$ and $|v\rangle$ with structure $|x^{*}\rangle\otimes|x\rangle$. Although this remains to be done, the approach presented herein shows the advantage of being easily extended to the multipartite case.

\section*{Acknowledgments}
We acknowledge financial support from Spanish Ministry of Education and Science through project No.\ FIS2005-01574.



\end{document}